\documentclass[12pt]{iopart}
\usepackage[latin1]{inputenc}
\usepackage{iopams}

\begin{document}
\title[${\cal N}=1$ SUSY Wong Equations and the Non-Abelian Landau Problem]{The ${\cal N}=1$ Supersymmetric Wong Equations and the Non-Abelian Landau Problem}

\author{Micha\"el Fanuel$^1$,  Jan Govaerts$^{1,2}$, Gabriel Y. H. Avossevou$^{3,4}$ and  Anselme F. Dossa$^{3}$}
\vspace{10pt}
\address{$^1$Centre for Cosmology, Particle Physics and Phenomenology (CP3),\\
Institut de Recherche en Math\'ematique et Physique,\\
Universit\'e catholique de Louvain, Chemin du Cyclotron 2, bte L7.01.01,\\
B-1348, Louvain-la-Neuve, Belgium}
\ead{Michael.Fanuel@uclouvain.be, Jan.Govaerts@uclouvain.be}
\vspace{10pt}
\address{$^2$International Chair in Mathematical Physics\\
and Applications (ICMPA--UNESCO Chair),\\
University of Abomey--Calavi, 072 B. P. 50, Cotonou, Republic of Benin}
\vspace{10pt}
\address{$^3$Unit\'e de Recherche en Physique Th\'eorique (URPT),\\
 Institut de Math\'ematiques et de Sciences Physiques (IMSP),\\
01 B. P.  613 Porto--Novo,  Republic of Benin}
\ead{gabavossevou@gmail.com, finedofas@yahoo.fr}
\vspace{10pt}
\address{$^4$D\'epartement de Physique,\\
 Universit\'e d'Abomey--Calavi (UAC),  Republic of Benin}

\begin{abstract}
A Lagrangian formulation is given extending to ${\cal N}=1$ supersymmetry the motion of a charged point particle with spin
in a non-abelian external field.  The classical formulation is constructed for any external static non-abelian SU(N) gauge potential.
As an illustration, a specific gauge is fixed enabling canonical quantization and the study of the supersymmetric
non-abelian Landau problem. The spectrum of the quantum Hamiltonian operator follows in accordance with the supersymmetric structure.
\end{abstract}

\section{Introduction}

The motion of a non-abelian charged particle in a classical non-abelian gauge field as described many years ago by
the Wong equations \cite{Wong} has been studied for its theoretical interest \cite{Balachandran,Wipf, Horvathy}
and in attempts to describe the phenomenology of the strong interactions \cite{Jeon}.
More recently a study of the non-abelian Landau problem \cite{Estienne}, {\it i.e.}, a quantum particle confined to a plane
and subjected to a static and homogeneous perpendicular magnetic field \cite{Landau}, has shown that the effects of specific choices
of non-abelian gauge potentials corresponding to homogeneous coloured magnetic fields could account for the presence of
Rashba \cite{Rashba1,Rashba2} and Dresselhaus \cite{Dresselhaus} spin-orbit interactions. Applications of similar models
have been recently analyzed in the context of Fractional Quantum Hall Effects \cite{Pachos}, and Non-Abelian Anyons \cite{Burrello1, Burrello2}.
These results have revived the interest in studying the motion of coloured particles in non-abelian backgrounds.
Undoubtedly the non-abelian Landau problem is a particular example of the motion of a coloured point particle in an external
Yang-Mills gauge field. At the classical level this motion is governed by the Wong equations, generalizing ordinary cyclotron motion
in the case of an electromagnetic field.

Inspired by the potential application of supersymmetry in condensed matter physics \cite{EzawaM}, the consequences of having in addition
a supersymmetric invariant realization of the quantized system corresponding to the motion of a coloured particle in a classical
external static non-abelian gauge field are studied in the present work. The authors of Ref.\cite{vanHolten} addressed, in a classical formulation,
the ${\cal N}=1$ supersymmetric Wong equations for a spin $1/2$ particle in an non-abelian background gauge field. The present work considers the case of a particle with arbitrary spin in a unitary (irreducible) representation of a compact gauge group. Furthermore, a canonical
quantization of the classical formulation is constructed. Subsequently, as a particular illustration, the spectrum of the ${\cal N}=1$
supersymmetric non-abelian Landau problem is obtained in the specific case of a spin $1/2$ particle in a non trivial static non-abelian background magnetic field. Finally, the inclusion of an electric potential term is discussed in the Appendix.

\section{Lagrangian and Supersymmetry}

A typical problem is the motion of a charged particle in $d$ spatial dimensions subjected to a generic static non-abelian gauge potential. Its classical dynamics is governed by
\begin{eqnarray}
L=\frac{1}{2}m\dot{x}_{i}^{2}+\frac{1}{2}\rmi(u_{\alpha}^{\dagger}\dot{u}_{\alpha}-\dot{u}_{\alpha}^{\dagger}u_{\alpha})+\dot{x}_{i}A_{i}^{a}(x_{i})u_{\alpha}^{\dagger}T_{\alpha\beta}^{a}u_{\beta}-A_{0}^{a}(x_{i})u_{\alpha}^{\dagger}T_{\alpha\beta}^{a}u_{\beta},\label{One}
\end{eqnarray}
where we have introduced additional complex variables $u_{\alpha}$ that will be specified hereafter while $A_{0}^{a}(x_{i})$ and $A_{i}^{a}(x_{i})$ are arbitrary time independent background fields ($i=1,\dots, d$).
For simplicity we take a compact gauge group, with hermitian generators, $(T^{a})^{\dagger}=T^{a}$, whose algebra is $[T^{a},T^{b}]=\rmi f^{abc}T^{c}$. By construction the Lagrangian is invariant under local space dependent gauge transformation, which includes a transformation of the background fields,
\begin{eqnarray}
&u_{\alpha}\to g(x_{i})_{\alpha\beta}u_{\beta},\\
&A_{i}^{a}(x_{i})T^{a}\to g(x_{i})A_{i}^{a}(x_{i})T^{a}g(x_{i})^{\dagger}-\rmi\partial_{i}g(x_{i}) g(x_{i})^{\dagger},\label{GaugeTransfBackground}\\
&A_{0}^{a}(x_{i})T^{a}\to g(x_{i})A_{0}^{a}(x_{i})T^{a}g(x_{i})^{\dagger},
\end{eqnarray}
where $g(x_{i})$ is a group element with space dependent parameters $\lambda^{a}(x_{i})$, obtained by exponentiation of the elements of the algebra: 
$g(x_{i})=\exp \rmi\lambda^{a}(x_{i})T^{a}$. In order to introduce the supersymmetric formulation we define the hermitian supercharge and superderivative 
\begin{eqnarray}
Q=\partial_{\theta}+\rmi\theta\partial_{t},\quad
D=\partial_{\theta}-\rmi\theta\partial_{t},
\end{eqnarray}
which anticommute with each other while $Q^{2}=\rmi\partial_{t}$ and $D^{2}=-\rmi\partial_{t}$. The supersymmetric formulation requires the introduction of the real Grassmann odd coordinate $\theta$. Furthermore it is customary to define the real supercoordinate $X_{i}(t,\theta)=x_{i}(t)+\rmi\theta\lambda_{i}(t)$ with $i=1, \dots, d$ where the variable $\lambda_{i}(t)$ is odd.
The infinitesimal supersymmetric transformation of $X_{i}$ of Grassmann odd parameter $\epsilon$ is
$\delta_{\epsilon}X_{i}(t,\theta)=-\rmi\epsilon Q X_{i}(t,\theta)$,
so that the components transform as
\begin{eqnarray}
\delta_{\epsilon}x_{i}(t)=\epsilon \lambda_{i}(t), \quad
\delta_{\epsilon}\lambda_{i}(t)=\rmi\epsilon \dot{x}_{i}(t).\label{SUSY1}
\end{eqnarray}
With the help of these definitions we readily find that the kinetic term $m\dot{x}_{i}^{2}/2$ is contained in the expression
\begin{eqnarray}
\int d\theta \{-\frac{1}{2}mD^{2}X_{i}DX_{i}\}=\frac{1}{2}m\dot{x}_{i}^{2}-\frac{\rmi}{2}m\dot{\lambda}_{i}\lambda_{i}.
\end{eqnarray}
By construction, the last expression transforms by a total derivative under an infinitesimal SUSY transformation. We briefly recall the reason why this is so.
As a general rule the Grassmann integral of a superfield $\mathbb{L}=L^{(1)}+\theta L^{(2)}$ selects
$\int d\theta \mathbb{L} =L^{(2)}$. The highest component $L^{(2)}$ has the property of transforming by a total derivative under supersymmetric transformations.
Indeed, the transformation
\begin{eqnarray}
\delta_{\epsilon}\mathbb{L}=-\rmi\epsilon L^{(2)}-\theta\epsilon\partial_{t}L^{(1)}\label{SUSYcharge}
\end{eqnarray}
is such that we can identify the transformation rule $\delta_{\epsilon}L^{(2)}=-\epsilon \partial_{t}L^{(1)}$.
This construction produces a natural candidate for the supersymmetric extension of the original Lagrangian.

\section{Bosonic Degree of Freedom}

Henceforth we will consider $u_{\alpha}$ to be a Grassmann even complex variable, in a (complex) representation of the ``gauge'' group. The case of a Grassmann odd variable can be developed in parallel with our analysis.
The complex supercoordinate containing $u_{\alpha}$ reads $U_{\alpha}(t,\theta)=u_{\alpha}(t)+\theta w_{\alpha}(t) $ whose components transform under supersymmetry as 
\begin{eqnarray}
\delta u=-\rmi\epsilon w,\quad \delta w=-\epsilon \dot{u}\label{SUSY3}.
\end{eqnarray}
\subsection{Non-Abelian Transformations and Supersymmetric Action}
Respecting gauge invariance requires particular care. Here we consider a background with vanishing electric potential while Appendix A shows how to generalize the formulation to include an electric potential, $A^{a}_{0}(x)\neq 0$.
Inspired by the supersymmetrization in the Abelian Landau problem and references \cite{rietdijk1, rietdijk2}, we define the natural candidate
$V=\rmi A(X_{\mu},DX_{\mu})=\rmi DX_{j}A^{a}_{j}(X_{i})T^{a}$, for the analogue of the vector superfield. The components of $V$ are
\begin{eqnarray}
V=-\lambda_{i}A_{i}(x_{i})+\theta\{\dot{x}_{i}A_{i}(x_{i})+\rmi\lambda_{i}\lambda_{j}\partial_{i}A_{j}(x_{i})\},
\end{eqnarray}
where $A_{i}(x_{i})=A_{i}^{a}(x_{i})T^{a}_{\alpha\beta}$ for $i,j=1,\dots, d$. Henceforth the gauge index of the representation is kept implicit. In this $0+1$ dimensional problem, there is no need to impose a ``Wess-Zumino'' gauge due to the small number of components.
However we still need to investigate the gauge transformation properties of the vector superfield in order to define an invariant action. 
In the case of a $U(1)$ gauge group \cite{BenGelounGovaertsScholtz}, the gauge transformation of the vector superfield is 
$V\to V-\rmi D\Lambda$
where the superfield $\Lambda$ is real and reads
 \begin{eqnarray}
\Lambda(X_{i})=\Lambda^{a}(x_{i})T^{a}+\rmi\theta\lambda_{i}\partial_{i}\Lambda^{a}(x_{i})T^{a}
\end{eqnarray}
so that
\begin{eqnarray}
\rmi D\Lambda(X_{i})=\{-\lambda_{i}\partial_{i}\Lambda^{a}(x_{i})+\theta \dot{x_{i}}\partial_{i}\Lambda^{a}(x_{i})\}T^{a}.
\end{eqnarray}
It is straightforward to extend the transformation for the case of the SU(N) gauge group. The non-abelian transformation of the vector superfield
\begin{eqnarray}
V\to e^{\rmi\Lambda}Ve^{-\rmi\Lambda}+e^{\rmi\Lambda}De^{-\rmi\Lambda}.
\end{eqnarray}
when written in its infinitesimal form, $V\to V+\delta V$, reads
\begin{eqnarray*}
\delta V&=-\rmi D\Lambda+[\rmi\Lambda;V]\\
&=\lambda_{i}(x_{i})(\partial_{i}\Lambda(x_{i})+f^{abc}\Lambda^{a}(x_{i})A^{b}_{i}(x_{i})T^{c})\\
&\quad -\theta\big\{\dot{x}_{i}(\partial_{i}\Lambda+f^{abc}\Lambda^{a}(x_{i})A^{b}_{i}(x_{i})T^{c})\\
&\quad +\rmi\lambda_{i}\lambda_{j}\partial_{j}(\partial_{i}\Lambda(x_{i})+f^{abc}\Lambda^{a}(x_{i})A^{b}_{i}(x_{i})T^{c})\big\},
\end{eqnarray*}
providing the desired result.
Besides, the gauge transformation of the variable $u$ is $u\to \exp(i\lambda^{a}(x_{i})T^{a})u$ and therefore the supersymmetric extension is easily guessed. The gauge transformation, of parameter $\Lambda(X)=\Lambda(X)^{\dagger}$ valued in the Lie algebra representation, of the superfieds $U$ and $U^{\dagger}$ is
\begin{eqnarray}
U\to e^{\rmi\Lambda}U, \quad U^{\dagger} \to U^{\dagger}e^{-\rmi\Lambda}.
\end{eqnarray}
The supersymmetric generalization of the third term of the Lagrangian \eref{One}
\begin{eqnarray}
L_{3}=\int d\theta U^{\dagger}(-D+V)U
\end{eqnarray}
is straightforwardly written in components
\begin{eqnarray}
L_{3}=&\{\rmi u^{\dagger}\dot{u}+u^{\dagger}\dot{x}_{i}A_{i}(x_{i})u+\rmi u^{\dagger}\lambda_{i}\lambda_{j}\partial_{j}A_{i}(x_{i})u\nonumber\\
&-u^{\dagger}A_{0}(x_{i})u+w^{\dagger}w+w^{\dagger}\lambda_{i}A_{i}(x_{i})u+u^{\dagger}\lambda_{i}A_{i}(x_{i})w\}.\label{Auxilliary}
\end{eqnarray}
Furthermore the complete expression whose first term provides the supersymmetric variation of $L_{3}$ is
$U^{\dagger}(-D+V)U=(-u^{\dagger}w-u^{\dagger}\lambda_{i}A_{i}u)+\theta L_{3}$,
and, according to \eref{SUSYcharge} , gives $\delta_{\epsilon}L_{3}=-\epsilon\partial_{t}(-u^{\dagger}w-u^{\dagger}\lambda_{i}A_{i}u)$.
At first sight, the last expression for the Lagrangian \eref{Auxilliary} is not manifestly gauge invariant. However the variables $w$ and $w^{\dagger}$ have no dynamics and hence are auxilliary. The elimination of these variables by their equations of motion gives
\begin{eqnarray}
L_{3}=&u^{\dagger}\big(\rmi\partial_{t}+\dot{x}_{i}A_{i}(x_{i})\big)u\nonumber\\
&+\frac{1}{2}\rmi u^{\dagger}\lambda_{i}\lambda_{j}\{\partial_{j}A_{i}(x_{i})-\partial_{i}A_{j}(x_{i})-\rmi [A_{j}(x_{i});A_{i}(x_{i})]\}u,
\end{eqnarray}
which is explicitly invariant under a non-abelian gauge transformation provided the background field $A_{i}(x_{i})$ is transformed as well according to \eref{GaugeTransfBackground}.
Finally we notice that the results obtained are not explicitly real but we provide a remedy by considering instead
 \begin{eqnarray}
 L_{3}=\int d\theta \frac{1}{2}\Big\{[(-D+V)U]^{\dagger}U+U^{\dagger}(-D+V)U\Big\}
 \end{eqnarray}
 so that the supersymmetric variation becomes $\delta_{\epsilon}L_{3}=-\epsilon\partial_{t}(-\frac{1}{2}u^{\dagger}w-\frac{1}{2}w^{\dagger}u-u^{\dagger}\lambda_{i}A_{i}u)$, vanishing when the equations of motion are satisfied.
In conclusion, the supersymmetric action 
\begin{eqnarray}
S=\int dt d\theta \{-\frac{1}{2}mD^{2}X_{i}DX_{i}-L_{3}\}=\int dt L_{SUSY}
\end{eqnarray}
reads in components
\begin{eqnarray}
S=\int dt\Big\{&\frac{1}{2}m\dot{x}_{i}^{2}-\frac{\rmi}{2}m\dot{\lambda}_{i}\lambda_{i}+\frac{\rmi}{2}u^{\dagger}\dot{u}-\frac{\rmi}{2}\dot{u}^{\dagger}u\nonumber\\
&+u^{\dagger}\dot{x}_{i}A_{i}(x_{i})u+\frac{1}{2}\rmi u^{\dagger}\lambda_{i}\lambda_{j}G_{ji}u\Big\}.\label{action}
\end{eqnarray}
The non-abelian field strength $G_{ji}=\partial_{j}A_{i}(x_{i})-\partial_{i}A_{j}(x_{i})-\rmi [A_{j}(x_{i});A_{i}(x_{i})]$ transforms under gauge transformation by a conjugation, so that $u^{\dagger}G_{ij}u$ a gauge invariant expression.
The variation of this Lagrangian under supersymmetry
\begin{eqnarray}
\delta_{\epsilon}L_{SUSY}=\partial_{t}\big(\frac{1}{2}m\dot{x}_{i}\lambda_{i}\big)
\end{eqnarray}
gives rise to a surface term in the action, when the equations of motion hold.
The supersymmetry charge is readily found to be $Q_{susy}=m\dot{x}_{i}\lambda_{i}$. 

\subsection{Hamiltonian Formulation}
The introduction of additional variables $p_{i}$ for $i=1, \dots, d$ provides an alternative formulation of the action \eref{action}
\begin{eqnarray}
S=\int dt &\{\dot{x}_{i}p_{i}-\frac{\rmi}{2}m\dot{\lambda}_{i}\lambda_{i}+\frac{\rmi}{2}u^{\dagger}\dot{u}-\frac{\rmi}{2}\dot{u}^{\dagger}u\nonumber\\
&-\frac{1}{2m}(p_{i}-u^{\dagger}A_{i}u)^{2}+\frac{\rmi}{2}\lambda_{i}\lambda_{j}u^{\dagger}G_{ij}u\}.
\end{eqnarray} 
The last action being already in the so-called first order Hamiltonian form we take advantage of its convenient form to read off the Hamiltonian
\begin{eqnarray}
H=\frac{1}{2m}(p_{i}-u^{\dagger}A_{i}u)^{2}+\frac{\rmi}{2}\lambda_{i}\lambda_{j}u^{\dagger}G_{ij}u
\end{eqnarray}
as well as the Grassmann graded Poisson (or Dirac)  brackets
\begin{eqnarray}
\{x_{i};p_{j}\}=\delta_{ij}, \quad \{u;u^{\dagger}\}_{D}=-\rmi, \quad
\{\lambda_{i};\lambda_{j}\}_{D}=-\frac{i}{m}\delta_{ij}.
\end{eqnarray}
A straightforward analysis of constraints following the Dirac algorithm \cite{GovaertsBook} reaches the same conclusions.
In the Hamiltonian formalism the supercharge $Q_{susy}=(p_{i}-u^{\dagger}A_{i}^{a}T^{a}u)\lambda_{i}$ generates the infinitesimal supersymmetry transformations through the Poisson  brackets.
The transformations
\begin{eqnarray}
&\{\epsilon Q_{susy};x_{j}\}=-\delta_{ij}\epsilon\lambda_{i}, \
\{\epsilon Q_{susy};\lambda_{j}\}=-\rmi\delta_{ij}\epsilon\dot{x}_{i},\\
&\{\epsilon Q_{susy};u\}=-\rmi\epsilon A_{i}^{a}(x_{i})T^{a}\lambda_{i}u=\epsilon w,
\end{eqnarray}
correspond up to a sign to the ones in \eref{SUSY1} and \eref{SUSY3}.
As is very well known the supercharge is a square root of the Hamiltonian:
$\{Q_{susy},Q_{susy} \}=-2\rmi H$, where the curly brackets denote Poisson brackets.

\section{Canonical Quantization}

In accordance with the correspondence principle we proceed to canonical quantization by introducing the following commutators and anti-commutators:
\begin{eqnarray}
[\hat{x}_{i};\hat{p}_{j}]=\rmi\hbar\delta_{ij},\quad [u;u^{\dagger}]=\hbar, \quad\{ \lambda_{i};\lambda_{j}\}=\frac{\hbar}{m}\delta_{ij}.
\end{eqnarray}
Henceforth all curly brackets $\{ \ ; \ \}$ will denote anticommutators. In addition we introduce conveniently ``normalized'' operators
$\lambda_{i}=\sqrt{\hbar/2m}\gamma_{i}$,
satisfying a Clifford algebra $\{\gamma_{i};\gamma_{j}\}=2\delta_{ij}$ in Euclidian signature. 
The algebra of the ``non-abelian charges'':
$[I^{a};I^{b}]=\rmi\hbar f^{abc}I^{c}$
, where $I^{a}=u^{\dagger}T^{a}u$, contains an additionnal factor $\hbar$ with respect to the Lie algebra considered at the beginning.
The Hamiltonian operator easily follows
\begin{eqnarray}
\hat{H}=\frac{1}{2m}(\hat{p}_{i}- A_{i}^{a}(\hat{x}_{i})I^{a})^{2}-\frac{\hbar}{4m}\rmi\gamma_{i}\gamma_{j}G_{ij}^{a}(\hat{x}_{i})I^{a}.
\end{eqnarray}
The convenient notation $\sigma_{ij}=\frac{\rmi}{2}[\gamma_{i};\gamma_{j}]$ renders manifest the fact that we deal with a spinor representation of \cal{SO}$(d)$.
The operator corresponding to the supercharge 
\begin{eqnarray}
\hat{Q}_{susy}=(\hat{p}_{i}-u^{\dagger}A_{i}^{a}(\hat{x}_{i})T^{a}u)\lambda_{i}/\sqrt{\hbar}
\end{eqnarray}
 is also the square root of $\hat{H}$, as readily shown after a short algebra:
$\hat{Q}^{2}_{susy}=\hat{H}$.
\subsection{Constant Non-Abelian Magnetic Field}
As an illustration we discuss the particular case corresponding to the problem of a particle confined in a plane in a non-abelian homogeneous magnetic field. 
For simplicity we restrict ourselves to the $U(1)\times SU(2)$ gauge group (with $a=0, 1, 2, 3$) and consider the fundamental representation of $SU(2)$.
 Representing the $\lambda_{i}$ sector with the help of Pauli matrices ($[\sigma_{1};\sigma_{2}]=2\rmi\sigma_{12}=2\rmi\sigma_{3}$), we find that the Hamiltonian takes the suggestive form
\begin{eqnarray}
\hat{H}=\frac{1}{2m}(\hat{p}_{i}- A_{i}^{a}(\hat{x}_{i})I^{a})^{2}+\frac{\hbar}{2m}\sigma_{3}G_{12}^{a}(\hat{x}_{i})I^{a}.
\end{eqnarray}
In order to study the corresponding Landau problem \cite{Estienne}, let us choose the non-abelian vector potential
$A_{i}(\hat{x}_{i})=-(\frac{1}{2}B\epsilon_{ij}\hat{x}_{j}+\beta\sum_{a=1}^{3}\epsilon_{ia}I^{a})$, leading to the non-abelian magnetic field
$G_{12}=B\mathbb{I}+\hbar\beta^{2}I^{3}$. Consider then the Fock operators (see for example \cite{Govaerts})
\begin{eqnarray}
a_{i}=\frac{1}{2}\sqrt{\frac{m\omega_{c}}{\hbar}}(\hat{x}_{i}+\frac{2\rmi}{m\omega_{c}}\hat{p}_{i}),\quad a^{\dagger}_{i}=\frac{1}{2}\sqrt{\frac{m\omega_{c}}{\hbar}}(\hat{x}_{i}-\frac{2\rmi}{m\omega_{c}}\hat{p}_{i}),
\end{eqnarray}
with the cyclotron frequency $\omega_{c}=B/m$.
We introduce the ``chiral'' oscillators $(a_{\pm},a^{\dagger}_{\pm})$ which verify the Fock algebra $[a_{\pm};a^{\dagger}_{\pm}]=1$, defined through the expressions
\begin{eqnarray*}
a_{\pm}=\frac{1}{\sqrt{2}}(a_{1}\mp \rmi a_{2}),\quad a^{\dagger}_{\pm}=\frac{1}{\sqrt{2}}(a^{\dagger}_{1}\pm \rmi a^{\dagger}_{2}).
\end{eqnarray*}
Therefore the phase space operators $(\hat{x}_{i},\hat{p}_{i})$ can be expressed in terms of $(a_{\pm},a^{\dagger}_{\pm})$.
These notations prove themselves to be appropriate in order to rewrite the quantum supercharge
\begin{eqnarray}
\sqrt{2m}\hat{Q}_{susy}=-i(\sqrt{2\hbar B}a^{\dagger}_{-}+2\beta\Sigma_{+})b+\rmi(\sqrt{2\hbar B}a_{-}+2\beta\Sigma_{-})b^{\dagger},
\end{eqnarray}
where we defined the ladder operators $\Sigma_{\pm}=(\hat{I}_{1}\pm \rmi\hat{I}_{2})/2$ satisfying $[\Sigma_{+};\Sigma_{-}]=\pm\hbar\Sigma_{3}$. The fermionic Fock operators used above are defined by 
$b=(\gamma_{1}-\rmi\gamma_{2})/2$ and
$b^{\dagger}=(\gamma_{1}+\rmi\gamma_{2})/2$ and satisfy $\{b;b^{\dagger}\}=1$. The particular representation of the $(\Sigma_{\pm},\Sigma_{3})$ algebra considered here is that of ``spin'' $1/2$: 
$\Sigma_{\pm}|\mp\rangle=\hbar|\pm\rangle$, and $\Sigma_{3}|\pm\rangle=\pm\hbar|\pm\rangle$.
The $2\times 2$ matrix representation of the fermionic Fock algebra enables to recover the structure of supersymmetric quantum mechanics
\begin{eqnarray*}
\hat{Q}_{susy}=
\left(\begin{array}{cc}
0 & A^{\dagger}\\
A & 0
\end{array}\right)
\end{eqnarray*}
where 
\begin{eqnarray}
A=\rmi(\sqrt{\frac{\hbar B}{m}}a_{-}+2\frac{\beta}{\sqrt{2m}}\Sigma_{-}), \ A^{\dagger}=-\rmi(\sqrt{\frac{\hbar B}{m}}a^{\dagger}_{-}+2\frac{\beta}{\sqrt{2m}}\Sigma_{+}).
\end{eqnarray}
As a result the quantum Hamiltonian 
\begin{eqnarray*}
\hat{H}=\left(\begin{array}{cc}
 A^{\dagger}A & 0\\
0 & A A^{\dagger}
\end{array}\right)=\left(\begin{array}{cc}
\hat{H}_{B} & 0\\
0 & \hat{H}_{F}
\end{array}\right)
\end{eqnarray*}
manifests a more tractable structure that we elaborate further. It is the customary to name the operators $ A^{\dagger}A$ and $A A^{\dagger}$ the bosonic and fermionic Hamiltonians, respectively.
Considering the eigenvalue problem for the bosonic and fermionic Hamiltonian, it is well known that an eigenstate $|\psi_{B}\rangle$ of non-vanishing eigenvalue for the bosonic Hamiltonian has a superpartner eigenstate of the fermionic Hamiltonian $|\psi_{F}\rangle=A|\psi_{B}\rangle$ associated to the same eigenvalue.
As a matter of fact the case of an eigenstate with vanishing eigenvalue needs to be considered separately.
The bosonic Hamiltonian $A^{\dagger}A$ possesses a zero-energy eigenstate provided by the condition $A|\psi_{0B}\rangle=0$:
\begin{eqnarray}
|\psi_{0B}\rangle=|n_{-}=0,n_{+},-\rangle
\end{eqnarray}
with $|n_{-}=0,n_{+},\pm \rangle=|n_{-}\rangle\otimes |n_{+}\rangle\otimes |\pm \rangle$ where $|n_{\pm}\rangle$ are normalised chiral Fock states. Nevertheless the fermionic Hamiltonian $A A^{\dagger}$ does not have a zero-energy eigenstate so that $|\psi_{0B}\rangle$ has no superpartner.
In consequence the reminder of the spectrum is found among the non vanishing eigenvalues of the bosonic Hamiltonian
\begin{eqnarray}
\hat{H}_{B}=\hbar\omega_{c}a^{\dagger}_{-}a_{-}+\frac{\beta}{m}\sqrt{2\hbar B}(a_{-}^{\dagger}\Sigma_{-}+a_{-}\Sigma_{+})+2\frac{\beta^{2}}{m}\Sigma_{+}\Sigma_{-}.
\end{eqnarray}
It is a straightforward exercise to obtain the non vanishing eigenvalues of $\hat{H}_{B}$:
\begin{eqnarray}
\hspace{-26pt}E^{B}_{(n_{-},\pm)}=\Big(\hbar\omega_{c}(n_{-}+\frac{1}{2})+\frac{\hbar^{2}\beta^{2}}{m}\Big)\Big\{1\pm\sqrt{1-\frac{n_{-}(n_{-}+1) }{(n_{-}+\frac{1}{2}+\frac{\hbar\beta^{2}}{ m\omega_{c}})^{2}}}\Big\},
\end{eqnarray}
for $n_{-}\in \mathbb{N}$, associated to the eigenstates
\begin{eqnarray}
\hspace{-20pt}&\frac{1}{N_{(n_{-},\pm)}}|E^{B}_{(n_{-},\pm)}\rangle\\
\hspace{-20pt}&=\sqrt{2\hbar B(n_{-}+1)}\frac{\hbar\beta}{m}|n_{-},n_{+},+\rangle+(E^{B}_{(n_{-},\pm)}-2\frac{\hbar^{2}\beta^{2}}{m}-\hbar\omega_{c}n_{-})|n_{-}+1,n_{+},-\rangle\nonumber
\end{eqnarray}
where $N_{(n_{-},\pm)}$ is an appropriate normalization. Finally the eigenvectors of $\hat{H}_{F}$ given by $A|E^{B}_{(n_{-},\pm)}\rangle$ correspond to the same non zero eigenvalues, as is expected because of supersymmetry.
Note how each of these energy eigenvalues are doubly infinitely degenerate, corresponding to the supersymmetric extension of the ordinary Landau levels, except for the lowest Landau level which is $\mathcal{N}=1$ supersymmetry invariant and of zero energy eigenvalue.
\section{Conclusion}
Given the present construction of the $\mathcal{N}=1$ supersymmetric generalization of the Wong equations, within the restriction to the $U(1)\times SU(2)$ gauge group we discussed the supersymmetric structure of the spectrum of a particular case: the non-abelian Landau problem. The relevance of this model to low dimensional materials exhibiting spin-orbit interactions may be investigated. Besides its phenomenological interest the Landau problem illustrates important modern mathematical developements, such as non-commutative geometry. As a perspective the projection of the dynamics on selected Landau levels and an investigation of the ensuing consequences of supersymmetry on the non-commutative planar euclidian geometry may be carried out.
\ack
The work of MF is supported by the National Fund for Scientific Research
(F.R.S.-FNRS, Belgium) through a ``Aspirant'' Research fellowship. This work is supported by the Belgian Federal Office
for Scientific, Technical and Cultural Affairs through the Interuniversity Attraction Pole P6/11.
\appendix
\section{Electric potential}
As a generalization, it is possible to include in the Lagrangian a scalar potential term: $-u^{\dagger}A^{a}_{0}(x)T^{a}u$, where we restrict ourselves to the case of a bosonic $u$. The chromo-electric potential $A_{0}$ is valued in the Lie algebra of the gauge group. The results obtained below can probably be found by considering the non relativistic limit of the model for a spinning particle in a background field with $\mathcal{D}=1$ supergravity, as studied in \cite{vanHoltenSUGRA}.
In the present work we obtain the supersymmetric action from the superspace formalism. To do so we define in addition the real Grassmann odd superfields $Y=y_{0}+\theta y_{1}$ and $\Lambda=\Lambda_{0}+\theta\Lambda_{1}$
where $y_{0}$ and $\Lambda_{0}$ are real fermions, and $y_{1}$, $\Lambda_{1}$ are real bosons. With the help of these supercoordinates, we build
\begin{eqnarray}
&\mathbb{L}_{0}&=-YU^{\dagger}A_{0}(X_{i})U+\Lambda(DY-1)\\
&&=\Lambda_{0}(y_{1}-1)-y_{0}u^{\dagger}A_{0}(x_{i})u+\theta\Big(\rmi\Lambda_{0}\dot{y}_{0}+\Lambda_{1}(y_{1}-1)-y_{1}u^{\dagger}A_{0}(x_{i})u\nonumber\\
&&\quad +y_{0}u^{\dagger}A_{0}(x_{i})w+w^{\dagger}A_{0}(x_{i})uy_{0}+\rmi u^{\dagger}y_{0}\lambda_{i}\partial_{i}A_{0}(x_{i})u\Big),
\end{eqnarray}
whose highest component $L_{0}=\int\rmd \theta \ \mathbb{L}_{0}$ contains the term $-u^{\dagger}A_{0}(x_{i})u$ after solving the equation of motion for $\Lambda_{1}$. The result is
\begin{eqnarray}
&L_{0}=&
\rmi\Lambda_{0}\dot{y}_{0}-u^{\dagger}A_{0}(x_{i})u+\rmi u^{\dagger}y_{0}\lambda_{i}\partial_{i}A_{0}(x_{i})u\nonumber \\
&&+u^{\dagger}y_{0}A_{0}(x_{i})w+w^{\dagger}y_{0}A_{0}(x_{i})u.
\end{eqnarray}
The role of the variable $\Lambda_{0}$ is now clear, it is the conjugate momentum of $y_{0}$. The complete Lagrangian obtained from the highest component of the superfield
\begin{eqnarray}
&-\frac{1}{2}mD^{2}X_{i}DX_{i}+\frac{1}{2}\Big\{[(-D+V)U]^{\dagger}U+U^{\dagger}(-D+V)U\Big\}\nonumber\\
&-YU^{\dagger}A_{0}(X_{i})U+\Lambda(DY-1)
\end{eqnarray}
where $V=\rmi(DX_{i}) A_{i}(X_{i})$, and takes the following form in components

\begin{eqnarray}
L_{T}=&\frac{1}{2}m\dot{x}_{i}^{2}-\frac{\rmi}{2}m\dot{\lambda}_{i}\lambda_{i}+\rmi\Lambda_{0}\dot{y}_{0}+u^{\dagger}(\rmi\partial_{t}+\dot{x}_{i}A_{i}(x_{i}))u -u^{\dagger}A_{0}(x_{i})u\nonumber\\
&+w^{\dagger}w+w^{\dagger}\lambda_{i}A_{i}(x_{i})u+u^{\dagger}\lambda_{i}A_{i}(x_{i})w+u^{\dagger}y_{0}A_{0}(x_{i})w+w^{\dagger}y_{0}A_{0}(x_{i})u\nonumber\\
&+\rmi u^{\dagger}y_{0}\lambda_{i}\partial_{i}A_{0}(x_{i})u+\rmi u^{\dagger}\lambda_{i}\lambda_{j}\partial_{j}A_{i}(x_{i})u.
\end{eqnarray}
In order to make the expression explicitly invariant under time independent gauge transformations we solve the equations of motions for $w$ and $w^{\dagger}$, and we obtain
\begin{eqnarray}
L_{T}=&\frac{1}{2}m\dot{x}_{i}^{2}-\frac{\rmi}{2}m\dot{\lambda}_{i}\lambda_{i}+i\Lambda_{0}\dot{y}_{0}+u^{\dagger}(\rmi\partial_{t}+\dot{x}_{i}A_{i}(x_{i}))u -u^{\dagger}A_{0}(x_{i})u\nonumber\\
&-\rmi y_{0}\lambda_{i}u^{\dagger}G_{0i}u-\frac{\rmi}{2} \lambda_{i}\lambda_{j}u^{\dagger}G_{ij}u,\label{LastLagrangian}
\end{eqnarray}
with the ``chromo-electric'' field $G_{0i}=-\partial_{i}A_{0}(x_{i})-\rmi [A_{0};A_{i}](x_{i})$.
Within the Hamiltonian analysis, the Dirac bracket $\{y_{0};\Lambda_{0}\}_{D}=-\rmi$ is straightforwardly obtained and the classical supercharge is then given by the expression
\begin{eqnarray}
Q_{T}=(p_{i}-u^{\dagger}A_{i}u)\lambda_{i}+\Lambda_{0}+y_{0}u^{\dagger}A_{0}u,
\end{eqnarray}
which verifies $\{Q_{T};Q_{T}\}=-2\rmi H_{T}$ with
\begin{eqnarray}
H_{T}=\frac{1}{2m}(p_{i}-u^{\dagger}A_{i}u)^{2}+\frac{\rmi}{2}\lambda_{i}\lambda_{j}u^{\dagger}G_{ij}u+\rmi y_{0}\lambda_{i}u^{\dagger}G_{0i}u+u^{\dagger} A_{0}u.
\end{eqnarray}

\end{document}